# Two-Step Modification of Phonon Mean Free Paths for Thermal Conductivity Predictions of Thin-Film-Based Nanostructures

Qing Hao[*], Yue Xiao, Sien Wang

Department of Aerospace and Mechanical Engineering, University of Arizona, Tucson, AZ 85721 U.S.A

[*]Corresponding authors: Qing Hao (qinghao@email.arizona.edu)

**Abstract**

As one simple metamaterial, nanopatterns are often fabricated across a thin film so that the thermal transport can be manipulated. The etched sidewalls for these nanostructures are usually rough due to surface defects introduced during the nanofabrication, whereas the top and bottom film surfaces are smoother. In existing analytical models, the contrast between these surfaces has not been addressed and all boundaries are assumed to be diffusive for phonon reflection. In this paper, a new two-step approach to address this issue is proposed for phonon transport modeling of general thin-film-based structures. In this approach, the effective in-plane phonon mean free paths ($\Lambda_{Film}$) are first modified from the bulk phonon MFPs to account for the influence of the top/bottom film surfaces, with possibly enhanced probability of specular phonon reflection at cryogenic temperatures. This $\Lambda_{Film}$ is further modified to include the scattering by etched sidewalls with almost completely diffusive phonon scattering. Such a two-step phonon mean free path modification yields almost identical results as frequency-dependent phonon Monte Carlo simulations for etched nanowires and representative nanoporous thin films. This simple yet accurate analytical model can be applied to general thin-film-based nanostructures to combine the phonon size effects along orthogonal directions.

**Nomenclature**

*Abbreviations*

| | |
|---|---|
| 2D | Two-dimensional |
| 3D | Three-dimensional |
| BTE | Boltzmann transport equation |
| MBL | Mean beam length |
| MC | Monte Carlo |
| MFP | Mean free path |
| RIE | Reactive Ion Etching |
| SOI | Silicon on insulator |

*Greek Symbols*

| | |
|---|---|
| $\eta$ | Average film-surface roughness (m) |
| $\theta$ | Included angle between the phonon traveling direction and the cross-plane direction (rad) |
| $\Lambda_{Bulk}$ | Bulk phonon mean free path (m) |
| $\Lambda_{eff}$ | Effective phonon mean free path (m) |
| $\Lambda_{Film}$ | Effective in-plane phonon mean free path along a solid thin film (m) |
| $\Lambda_{Pore}{}^{*}$ | Dimensionless ratio $L_{eff}/p$ |
| $\lambda$ | Phonon wavelength (m) |
| $\phi$ | Porosity |
| $\omega$ | Phonon angular frequency (rad/s) |
| $\omega_{max,i}$ | Maximum phonon angular frequency for branch $i$ (rad/s) |

*Roman Symbols*

| | |
|---|---|
| $A$ | Pore surface area (m²) |
| $b$ | Nanoslot depth (m) |
| $c_p$ | Volumetric phonon specific heat (J/m³·K) |
| $D$ | Side length of a square nanowire (m) |
| $d_{wire}$ | Diameter of a circular nanowire (m) |
| $F$ | Heat conduction reduction due to porosity |
| $f$ | Differential phonon mean free path distribution for a bulk material (W/m²·K) |
| $h$ | Film thickness (m) |
| $k$ | Thermal conductivity (W/m·K) |
| $k_L$ | Lattice thermal conductivity (W/m·K) |
| $k_\parallel$ | In-plane lattice thermal conductivity along the $l$ direction (W/m·K) |
| $k_\perp$ | In-plane lattice thermal conductivity along the $p$ direction (W/m·K) |
| $L$ | Characteristic length of a nanostructure, as determined by the geometry (m) |
| $L_{eff}$ | Effective characteristic length of a nanostructure (m) |
| $l$ | Periodic length of a periodic nanoslot pattern (m) |
| $i$ | Phonon branch index |
| $P_1$ | Specularity for film-surface phonon reflection |
| $P_2$ | Specularity for phonon reflection at etched sidewalls |
| $p$ | Pitch of a nanoporous film with circular pores (m) |
| $p_s$ | Width of one period for a periodic nanoslot pattern (m) |
| $q$ | Wave vector (1/m) |
| $q_{max}$ | Cutoff wave vector (1/m) |
| $r$ | Thickness-width ratio of a rectangular nanowire |
| $S$ | Phonon suppression function |
| $V_{\text{Solid}}$ | Solid region volume (m³) |
| $v_{g,i}$ | Phonon group velocity for branch $i$ (m/s) |
| $w$ | Width of a rectangular nanowire (m) |
| $w_s$ | Neck width between adjacent periodic nanoslots (m) |

## 1. Introduction

In solid thin films, the lattice thermal conductivity $k_L$ and thus thermal conductivity $k$ can be dramatically reduced from that for the bulk counterpart, resulting from the film surface scattering of particle-like phonons.[1] Such classical phonon size effects have been widely studied for thermal management of electronic devices based on thin films.[2-4] In more recent studies, various porous patterns have been introduced to thin films to tune the in-plane $k_L$, such as periodic micro- to nano-pores,[5-13] periodic nanoslots,[14] and nanoladders.[15] When electrons with a typically much shorter mean free path (MFP) are less affected for their transport, the thermoelectric performance can be enhanced. In experiments, nanowires fabricated from a thin film using reactive ion etching (RIE) are often used to calibrate the thermal measurements and are compared with the nanoporous thin films.[9] Compared with nanowires synthesized by the vapor-liquid-solid method, dry-etched nanowires have very rough sidewalls and the effective wire width can be smaller due to surface defects. As the result, the RIE-etched nanowires[16] exhibit a much lower $k$ than synthesized nanowires.[17] Similar arguments can be found for nanoporous thin films etched with RIE or deep RIE, where the pore-edge defects can expand the effective pore diameter.[18, 19] In modeling, the dry-etched nanowires have a rectangular cross section, with relatively smooth top/bottom film surfaces and rough wire sidewalls. At cryogenic temperatures, the top and bottom film surfaces are expected to have enhanced specular phonon refection as known for a solid film,[1] whereas the wire sidewalls still have mostly diffusive phonon reflection. The contrast between the film surfaces and nanowire sidewalls cannot be easily handled in analytical modeling using a temperature-independent characteristic length for the whole structure. More broadly, the same issue also exists for nanoporous thin films at cryogenic temperatures. A simple method is thus required to incorporate the phonon size effects due to both the film thickness and the pore-edge phonon scattering.

In this work, a simple but accurate analytical model is developed, which extends our previous work on nanoporous films[14, 20] to dry-etched nanowires and general thin-film-based nanoporous structures. The model predictions agree well with frequency-dependent phonon Monte Carlo (MC) simulations that incorporate the specularity of individual surfaces and the exact three-dimensional structures. The proposed model is based on the two-step modification of the bulk phonon MFP ($\Lambda_{Bulk}$), which well addresses the phonon MFP reduction along the the cross-plane direction and in-plane direction. This model can be named as a "two-step model" based on the strategy for the phonon MFP modification.

## 2. Analytical model and phonon MC simulations

A rectangular nanowire, with its height $h$ as the film thickness and width $w$ defined by nanofabrication, can be viewed as the interception between two perpendicular thin films (Fig. 1a). For nanoporous thin films, they can be viewed as a two-dimensional (2D) porous structure intercepting with a thin film (Fig. 1b). In general, the influence of the top/bottom surfaces of a film is included by modifying the bulk phonon MFP $\Lambda_{Bulk}$. This modified phonon MFP is further modified to incorporate the boundary scattering by etched sidewalls of nanowires or nanopore/nanoslot edges.

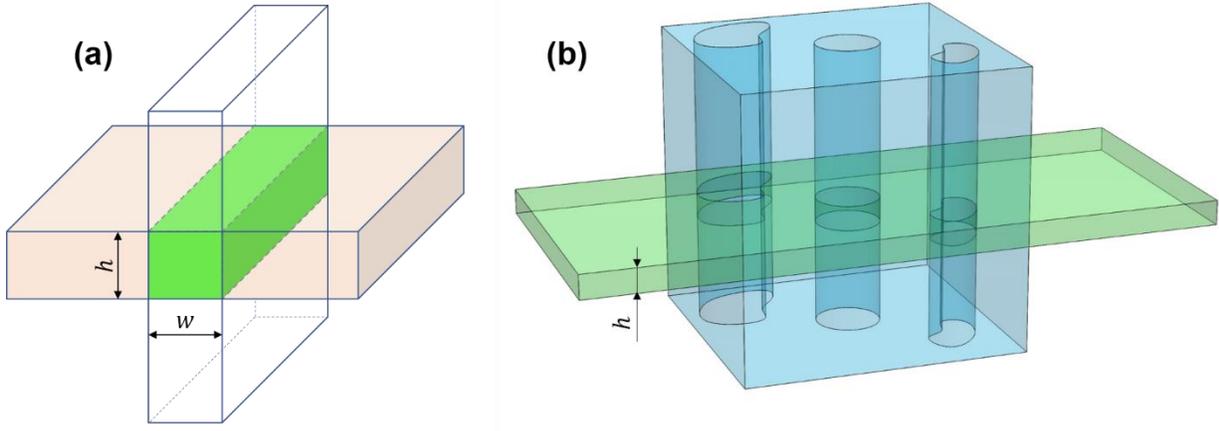

Fig. 1. (a) A rectangular nanowire (green) shown as the interception between two perpendicular thin films. In real samples, $h$ is the initial solid film thickness, while $w$ is defined by the nanofabrication. (b) A nanoporous film viewed as the interception between a thin film with its thickness $h$ (green) and a 2D nanoporous pattern (blue).

*2.1. Rectangular nanowires*

For rectangular nanowires, the two-step modification follows the Fuchs-Sondheimer model. First, the in-plane phonon MFP within a thin film ($\Lambda_{Film}$) is computed as[21]

$$\frac{\Lambda_{Film}}{\Lambda_{Bulk}} = 1 - \frac{3[1-P_1(\lambda)]\Lambda_{Bulk}}{2h} \int_0^1 (x - x^3) \frac{1-\exp\left(-\frac{h}{\Lambda_{Bulk}x}\right)}{1-P_1(\lambda)\exp\left(-\frac{h}{\Lambda_{Bulk}x}\right)} dx, \qquad (1)$$

where $P_1(\lambda)$ is the specularity of film-surface phonon reflection for a film with its thickness $h$. In Ziman's theory, the probability $P_1(\lambda)$ for specular reflection can be estimated as $P_1(\lambda) = exp\left[-16\pi^2 \left(\frac{\eta_1}{\lambda}\right)^2\right] = exp[-(2q\eta_1)^2]$, in which $\eta_1$ is the average film-surface roughness, $q$ is the wave vector, and $\lambda$ is phonon wavelength.[22] In the literature, $\pi^2$ is often wrongly replaced by $\pi^3$ and is corrected by Zhang and Lee.[23] A better understanding for film-surface phonon scattering can also be found in experiments[24-26] and simulations.[27-29] In this modification, a thin film is converted into another film with a reduced phonon MFP $\Lambda_{Film}$ but specular phonon reflection at film surfaces.

In the second step, the computed $\Lambda_{Film}$ is further modified as the effective phonon MFP $\Lambda_{eff}$ in a nanowire, given as

$$\frac{\Lambda_{eff}}{\Lambda_{Film}} = 1 - \frac{3[1-P_2(\lambda)]\Lambda_{Film}}{2w} \int_0^1 (x-x^3) \frac{1-\exp\left(-\frac{w}{\Lambda_{Film}x}\right)}{1-P_2(\lambda)\exp\left(-\frac{w}{\Lambda_{Film}x}\right)} dx, \qquad (2)$$

where $P_2(\lambda)$ is the phonon specularity for a film with its thickness $w$. In real samples, this width $w$ is defined by the nanofabrication. Depending on the corresponding surface roughness, $P_2(\lambda)$ can be very different from $P_1(\lambda)$. The modified $\Lambda_{eff}$ can then be used for the prediction of the axial lattice thermal conductivity $k_L$ based on the kinetic relationship:

$$k_L = \frac{1}{3}\sum_{i=1}^{3} \int_0^{\omega_{max,i}} c_p(\omega) v_{g,i}(\omega) \Lambda_{eff,i}(\omega) d\omega, \qquad (3)$$

where $c_p(\omega)$ and $v_{g,i}(\omega)$ are the differential volumetric phonon specific heat and phonon group velocity for the branch $i$ and angular frequency $\omega$, respectively. Without phonon confinement within ultrafine nanostructures to modify the phonon dispersion,[30, 31] both $c_p(\omega)$ and $v_{g,i}(\omega)$ are the same as those for bulk materials. Above procedure can be easily extended to general thermal analysis using the exact phonon dispersions and first-principles-computed phonon MFPs.[32] The energy-related integration in Eq. (3) is simply changed to the summation over all the phonon modes in the first Brillouin zone.

*2.2. General nanoporous thin films*

In a simple treatment for nanoporous structures, Eq. (2) becomes $\Lambda_{eff} = \left(1/\Lambda_{Film} + 1/L_{eff}\right)^{-1}$ based on the Matthiessen rule. Here $L_{eff}$ is the characteristic length of a 2D nanoporous thin film, i.e., a thin film with specular phonon reflection on its top/bottom surfaces.

This $L_{eff}$ is unchanged for films with different thicknesses and varied phonon specularity at the top/bottom film surfaces. The in-plane $k_L$ is now given as

$$k_L = \frac{F(\phi)}{3} \sum_{i=1}^{3} \int_0^{\omega_{max,i}} c_p(\omega) v_{g,i}(\omega) \Lambda_{eff,i}(\omega) d\omega, \tag{4}$$

in which $F(\phi)$ accounts for the heat transfer reduction due to the porosity $\phi$. In general, $F(\phi)$ can be determined by the Fourier's law analysis, e.g., as the thermal conductance ratio between a porous film and its nonporous counterpart.[32, 33] For periodic 2D nanoporous patterns, the Hashin-Shtrikman factor[34] is used and is given as

$$F(\phi) = \frac{1-\phi}{1+\phi}. \tag{5}$$

*2.3. Phonon MC simulations*

In this work, the analytical model is validated by comparing its predictions to phonon MC simulations that track the movement and scattering of individual phonons.[35] The solution for the phonon Boltzmann transport equation (BTE) can be statistically obtained when the code converges. As one big advantage, the exact structure geometry and frequency-dependent phonon MFPs can be both considered in these simulations. Using a boundary condition assuming a periodic heat flux and a constant virtual wall temperature,[35] the selected length of the simulated nanowire does not affect the result. For periodic nanoporous structures, a single period can be employed as the computational domain. A variance-reduced MC technique developed by Péraud and Hadjiconstantinou has also been employed to improve the computational efficiency.[36]

As the input of the MC simulations, the frequency-dependent bulk phonon MFPs are obtained by fitting the bulk-Si $k_L$, in which impurity scattering and Umklapp scattering are considered.[37] Only three identical sine-shaped acoustic branches are considered and the phonon dispersion is assumed to be isotropic. When phonon specularity should be calculated, the wave vector $q$ is computed with $\omega = \omega_{max} \sin\left(\frac{\pi}{2}\frac{q}{q_{max}}\right)$ for any given $\omega$, i.e., $q = 2q_{max} \arcsin(\omega/\omega_{max})/\pi$. Here $q_{max}$ is the cutoff wave vector at the edge of the first Brillouin zone. The corresponding $\lambda = 2\pi/q$ is used in Ziman's theory to compute $P(\lambda)$.

## 3. Results and discussion

Various thin-film-based nanostructures are studied with the proposed analytical model that is further validated with frequency-dependent phonon MC simulations. As one big improvement from previous studies, the partially specular and partially diffusive phonon reflection by the top/bottom film surfaces can be considered, instead of assuming complete diffusive phonon reflection by all boundaries. The discussions are focused on representative thin-film-fabricated structures, i.e., nanowires and periodic porous films, but it can be further extended to other geometries, including nanoladders,[15] fishbone nanowires,[38] films with Pacman pores,[39] and films with dog-leg-shaped pores.[40] The only requirement is to find out the characteristic length $L_{eff}$ for the 2D structure that intercepts with a thin film to yield the computed three-dimensional structure.

### 3.1. Rectangular nanowires

For a circular nanowire, the characteristic length $L$ as the nanowire diameter $d_{wire}$ is used to modify the bulk phonon MFP $\Lambda_{Bulk}$, i.e., $\Lambda_{eff} = (1/\Lambda_{Bulk} + 1/d_{wire})^{-1}$. This effective phonon MFP $\Lambda_{eff}$ is then used in Eq. (3) for thermal conductivity predictions. For a square nanowire, $L = 1.12D$ is used, with $D$ as the side of the square cross section.[41] Assuming diffusive nanowire boundary scattering of phonons, $L$ expression is given for a rectangular nanowire with its height $h$ and width $w$:[42]

$$L = \frac{\sqrt{hw}}{4}\left[3r^{-\frac{1}{2}}\ln(r^{-1}+\sqrt{1+r^{-2}}) + 3r^{\frac{1}{2}}\ln(r+\sqrt{1+r^2}) - r^{\frac{1}{2}}\sqrt{1+r^2} - r^{-\frac{1}{2}}\sqrt{1+r^{-2}} + r^{\frac{3}{2}} + r^{-\frac{3}{2}}\right], \tag{6}$$

in which the ratio $r = h/w$. Simplified expressions are also available. For instance, the square nanowire $L$ expression can be borrowed so that $L = 2\sqrt{hw/\pi} \approx 1.12\sqrt{hw}$ is proposed.[16, 43, 44] For $r > 1$ and better $r > 1.5$, Lee *et al.* also suggests $L = 3w(\ln 2r + 1/3r + 1/2)/4$ as an approximation for Eq. (6).[6]

Other than $L$ to modify the phonon MFPs, a suppression function $S(\Lambda_{Bulk})$ related to $h$ and $w$ can also be used to compute the lattice thermal conductivity:[45]

$$k_L = \int_0^\infty S(\Lambda_{Bulk}) f(\Lambda_{Bulk}) d\Lambda_{Bulk}, \tag{7}$$

where $f$ is the differential MFP distribution for the bulk material. The $S(\Lambda_{Bulk})$ expression for a rectangular nanowire is available.[46] In general, the above computational methods are only applicable when all nanowire boundaries are rough so that diffusive phonon boundary scattering is dominant.

In this work, nanowires with varied $h$ and $w$ values are computed for their lattice thermal conductivities. Assuming diffusive phonon reflection by all boundaries, Fig. 2a compares the room-temperature calculations using the procedure in Section 2.1 and those predicted by frequency-dependent phonon MC simulations. In comparison, Eq. (6) is used to calculate $k_L$ and the predictions are plotted as dashed lines. For a small $w/h$ ratio, it is found that $k_L$ predicted using Eq. (6) can be lower than the other predictions.

As an advancement, the proposed two-step model can be extended to a film with smoother top/bottom surfaces but two rough nanowire sidewalls etched by RIE. For a typical silicon on insulator (SOI) wafer, the surface roughness is 0.2–1 nm.[47] The two surfaces related to the film thickness $h$ is assigned with a 0.2 nm roughness as an extreme case here, whereas the other two etched sidewalls diffusively scatter all phonons. The simulations are carried out at 77 K to induce more specular phonon reflection at the top/bottom film surfaces. Figure 2b compares the $k_L$ yielded by phonon MC simulations (red solid line) and the analytical model (red triangle). The case with diffusive top/bottom film surfaces is also shown as blue circles (simulations) and dashed blue line (model). In general, the analytical model agrees well with the predictions by complicated phonon MC simulations.

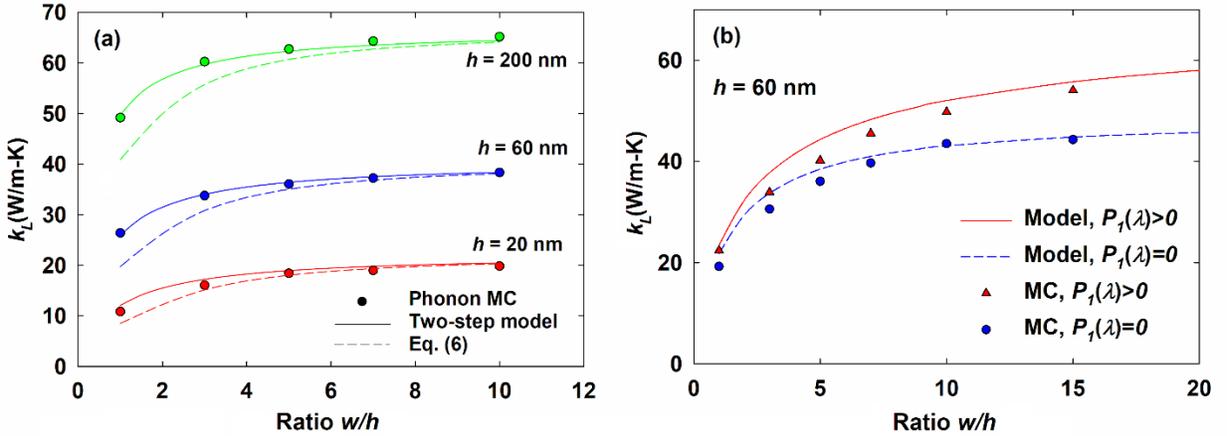

Fig. 2. Lattice thermal conductivities of a rectangular nanowire with (a) all diffusive surfaces at 300 K, and (b) partially specular phonon reflection by the top/bottom film surfaces at 77 K. All results from phonon MC simulations are in symbols and model predictions are in lines. The predictions using Eq. (6) is further plotted as dashed lines in (a) for comparison.

To further verify the two-step model, the experimental data on RIE-etched nanobeam as a rectangular nanowire, as given in Park et al.,[45] are compared with the model predictions. The film thickness $h$ is fixed at 78 nm, with a varied nanobeam width $w$. These nanowires are fabricated from the same Si film as the microdevice, which eliminates the critical thermal contact resistance between the measured nanostructure and the microdevice.[48, 49] With negligible electron contributions, the measured thermal conductivity $k$ is very close to the lattice part $k_L$. Figure 3 compares the model predictions (solid line) with measurement results at 300 K (symbols). All boundaries are assumed to be diffusive here. A dashed line predicted with Eq. (7) is extracted from Park et al. for comparison purposes. To be consistent with Park et al., first-principles phonon MFPs for bulk Si[50] are used in our calculations for Fig. 3. In general, two predictions generate very similar results, and both agree very well the experimental results.

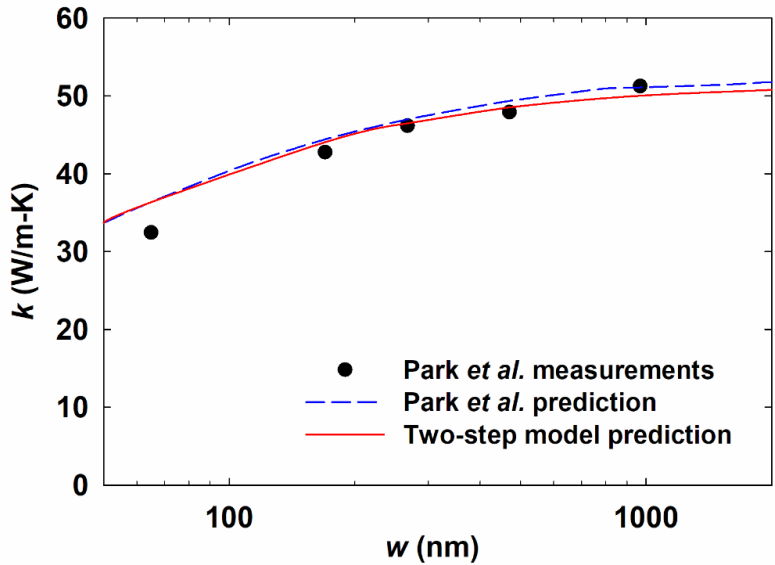

Fig. 3. Room-temperature thermal conductivity ($k \approx k_L$) of a nanobeam, with comparison between predictions using the two-step phonon MFP modification (solid line) and measurement results[45] (symbols). The predictions using a suppression function[45] is plotted as a dashed line.

*3.2. Films with periodic circular nanopores*

As one big advantage of the two-step model, this model can be easily extended to general nanoporous films. The in-plane phonon MFP $\Lambda_{Film}$ introduced by Eq. (1) can account for the partially or completely diffusive phonon scattering by the top/bottom surfaces of a film. With this conversion, a solid film can be viewed as a film with smooth top/bottom surfaces but an effective phonon MFP $\Lambda_{Film}$ reduced from the bulk $\Lambda_{Bulk}$. For this 2D structure, the nanopore influence can be further included using the characteristic length $L$. Restricted only by the 2D porous patterns, this $L$ is unchanged for different film thicknesses. For a thin film with periodic circular pores, various analytical expressions of $L$ have been proposed, as a function of the pitch and pore diameter.[51-54] However, some divergence can often be found between the model predictions and the predictions based on the phonon BTE. Detail discussions can be found in a recent review.[20] Along another line, an effective characteristic length $L_{eff}$ can be obtained by matching the in-plane $k_L$ predicted by Eq. (4) with that predicted by the frequency-dependent phonon MC simulations (Fig. 4a).[33] Here dimensionless $\Lambda_{Pore}^* = L_{eff}/p$ varies slightly for a changed pitch $p$. The plotted $\Lambda_{Pore}^*$ is thus averaged over those for the three pitch values, given as $p$=50, 200 and 500 nm. The given $L_{eff} = p \times \Lambda_{Pore}^*$ is anticipated to be accurate for general Si thin films with pitches of 50–500 nm. Figure 4b presents the predicted in-plane $k_L$ using $L_{eff}$ for the analytical model (lines), in comparison to that predicted by frequency-dependent phonon MC simulations (symbols).[19] The film thickness is fixed at 220 nm for all computed cases and all boundaries are assumed to be diffusive. In general, the model predictions agree with the predictions by the phonon MC simulations.

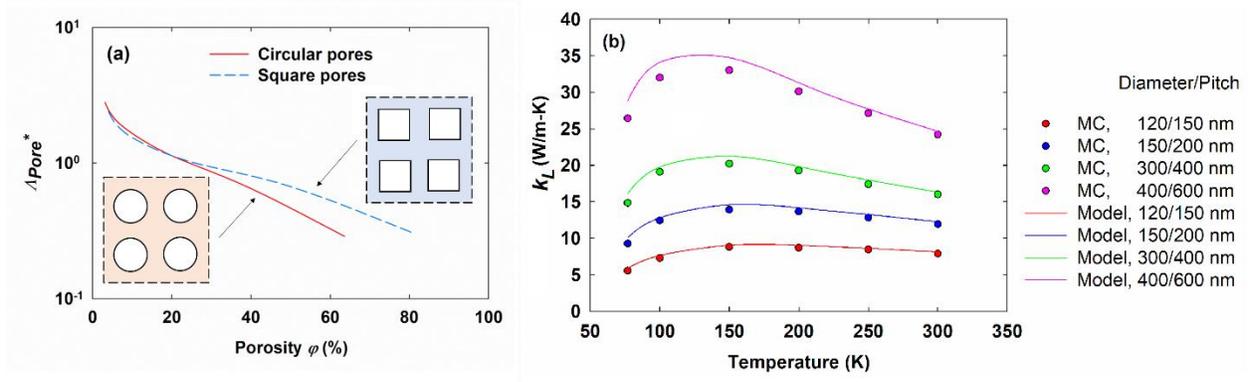

Figure 4. (a) Porosity- and period-dependent dimensionless $\Lambda_{Pore}^* = L_{eff}/p$, as summarized from figures in Hao *et al.*[33] (b) Comparison between phonon MC simulations[19] and the analytical model for temperature-dependent $k_L$ along the in-plane direction.

When $L_{eff}$ is unavailable, the radiative mean beam length (MBL) can be used as an estimated characteristic length, defined as MBL= $4V_{Solid}/A$.[33, 55] In its definition, the solid-region volume $V_{Solid}$ and pore surface area $A$ are evaluated within a single period. In practice, the MBL gives identical results as MC ray tracing under ballistic phonon transport.[33, 55] This treatment often gives some errors in the computed $k_L$ because it is inaccurate to use the Matthiessen rule to combine the boundary phonon scattering and internal phonon scattering inside the volume.[21] Different from a structure-determined characteristic length such as the MBL, the fitted $L_{eff}$ in Fig. 4a further depends on the phonon MFP distribution within a given material. The error associated with the Matthiessen rule is eliminated by directly matching the model predictions with the predictions by the phonon MC simulations. Similar effective $L_{eff}$ values have also been extracted in another study.[54]

*3.3. Films with periodic nanoslots*

The analytical model can also be extended to other nanoporous patterns fabricated across a thin film. For instance, $L_{eff}$ can be accurately predicted for a 2D structure with a row of periodic nanoslots drilled in the middle of the film.[14] With this $L_{eff}$ expression and $\Lambda_{Film}$ in Eq. (1), the in-plane $k_L$ can be predicted.

As a more general case, the anisotropic in-plane $k_L$ of a thin film with period nanoslots (Fig. 5) can also be predicted. When the $L_{eff}$ expression is unavailable, simulations of 2D porous pattern can be carried out first and compared to the prediction by Eq. (4), where $L_{eff}$ can be found by matching the two predictions. This $L_{eff}$ can then be used for films with different thicknesses and an arbitrary phonon specularity on the top/bottom film surfaces.

Without losing the generality, thin films with periodic nanoslots (Fig. 5a as the top view) are considered. The nanoslot pattern can effectively block the heat transport along the $l$ or "∥" direction by only allowing the phonons to travel through a channel with a narrow neck width $w_s$ and depth $b$. In contrast, the heat transport along the $p_s$ or "⊥" direction is largely intact. As a result, the nanoslot design leads to a large contrast between $k_\parallel$ and $k_\perp$, as the in-plane lattice thermal conductivities along the two major axis directions. Figures 5b and 5c show the temperature profiles of a representative 2D periodic nanoslot pattern with the heat flowing along the two directions. Around room temperature, an ~8 K temperature difference is applied across the simulated period with a periodic heat flux and constant virtual wall temperature boundary condition.[35] The employed dimensions are $w_s$=30 nm, $b$=5 nm, and $p_s$=$l$=100 nm. In Fig. 5b, a large temperature gradient can be found when phonons pass through the narrow neck. In Fig. 5c, the temperature profile generally has a more uniform gradient, except for some disturbance at the leading and back edges of the nanoslot. As anticipated, Fig. 5b usually leads to a much lower in-plane $k_L$.

Based on the 2D simulations, $L_{eff}$ used for Eq. (4) can be determined for $k_{\parallel}$ or $k_{\perp}$, respectively. Table I summarizes the $L_{eff}$ values for three representative nanoslot patters and the two heat-flow directions. These $L_{eff}$ values are then used for three-dimensional (3D) thin films with different thicknesses and diffusive film-surface phonon scattering. The anisotropic in-plane $k_L$ are shown in Fig. 5d, where the analytical model again agrees well with frequency-dependent phonon MC simulations.

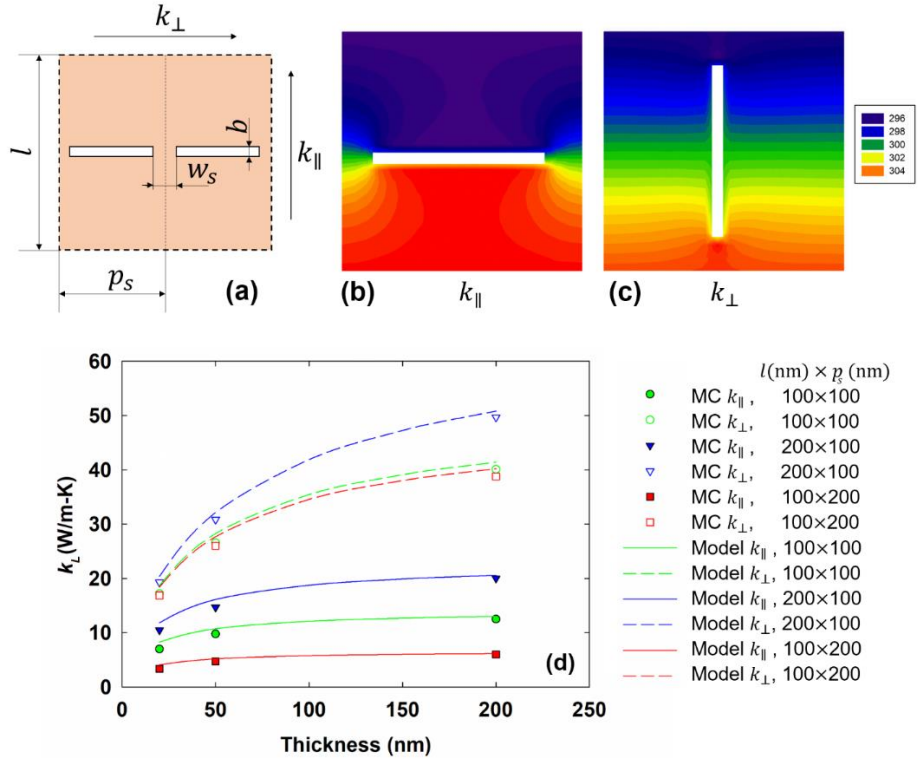

Fig. 5. (a) Top view of a thin film with patterned periodic nanoslots, with all geometry parameters marked. Two periods are shown here. Temperature profiles within a single period, with the heat flow along the (b) $l$ direction for $k_{\parallel}$ calculations, and (c) $p$ direction for $k_{\perp}$ calculations. In simulations for (b) and (c), the used dimensions are $w_s$=30 nm, $b$=5 nm, and $p_s$=$l$=100 nm. (d) Anisotropic in-plane $k_L$ of representative Si thin films with periodic nanoslots. All computations are at 300 K.

Table I. $L_{eff}$ values given by matching the predictions from frequency-dependent phonon MC simulations and Eq. (4) for different 2D nanoslot patterns.

| $w_s$ (nm) | $b$ (nm) | $l$ (nm) | $p_s$ (nm) | Direction | In-plane $k_L$ (W/m·K) | $L_{eff}$ (nm) |
|---|---|---|---|---|---|---|
| 30 | 5 | 100 | 100 | ∥ | 14.1 | 21.3 |
| | | | | ⊥ | 50.9 | 108.4 |
| | | 200 | 100 | ∥ | 22.9 | 34.0 |
| | | | | ⊥ | 68.1 | 224.8 |
| | | 100 | 200 | ∥ | 6.6 | 17.2 |
| | | | | ⊥ | 49.1 | 100.5 |

*3.4. Inverse thermal design of a nanoporous thin film*

With the given $\Lambda_{Bulk}$ of a material, the nanoporous pattern can be designed to modify the in-plane $k_L$ of a thin film for specified requirements. For a Si film with aligned circular nanopores, Fig. 6 displays the dependence of the in-plane $k_L$ on the pitch $p$ and porosity $\phi$, using $F(\phi)$ in Eq. (5) and $L_{eff}$ in Fig. 4a. The film thickness is fixed as 100 nm and all boundaries are assumed to be diffusive at 300 K. In this $k_L$ contour, the possible combination of $p$ and $\phi$ for a targeted in-plane $k_L$ can be easily found. In practice, the selection can further incorporate factors such as the spatial resolution of the nanofabrication and mechanical strength of the nanoporous films.

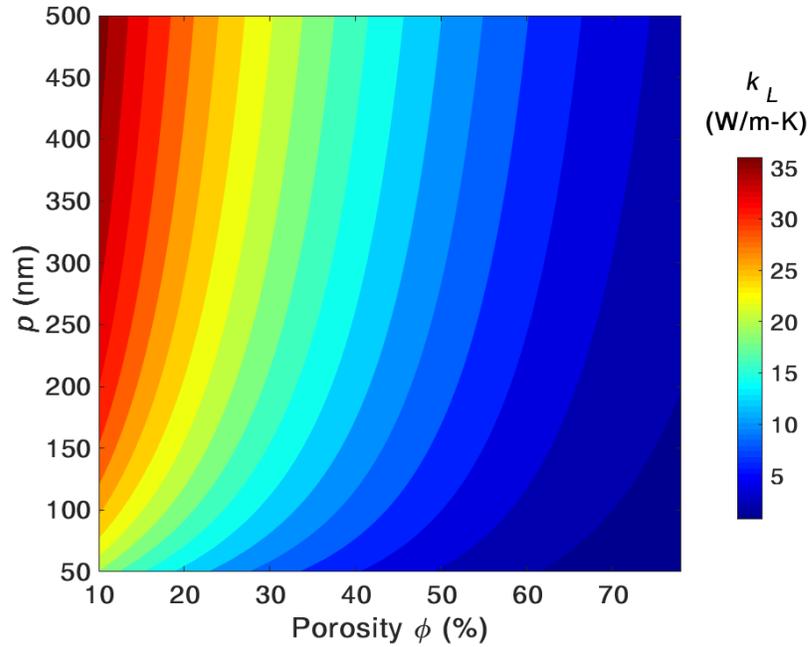

Fig. 6. In-plane $k_L$ of a 100-nm-thick Si thin film with varied pitches and porosities. The calculation is at 300 K and all boundaries are diffusive for phonon reflection.

### 3.5. Comparison to other techniques used for a nanoporous thin film

In existing studies, three-dimensional porous films have been calculated with multiple techniques. Accurate $k_L$ predictions can be provided by frequency-dependent phonon MC simulations,[33, 35, 54] as an alternative to directly solving the phonon BTE. However, such simulations are usually complicated and other techniques are thus pursued. With MC ray tracing or path sampling, the effective phonon MFP $\Lambda_{eff}$ in Eq. (4) can be obtained for a given $\Lambda_{Bulk}$ and an arbitrary three-dimensional structure.[6, 56-58] The Matthiessen rule may not be used here. The advantage of this MC-based $\Lambda_{eff}$ determination lies in that the $k_L$ calculations can easily incorporate the exact phonon dispersion and wave-vector-dependent bulk phonon MFPs.[32] However, such calculations should be carried out across the whole phonon MFP spectrum and must be completely repeated for varied film thicknesses, which is computationally expensive in many cases.

Instead of modifying the phonon MFPs, the suppression function in Eq. (7) can be obtained by solving a MFP-dependent phonon BTE for phonon transport along a nanoporous film.[59, 60] In general, numerically solving the phonon BTE is complicated with the discretization of the solid angle. In practice, it is computationally heavy to repeatedly carry out such calculations for varied film thicknesses and phonon specularity at the film surfaces.

In the analytical model, a travel-direction-dependent characteristic length $h/2|cos\theta|$ related to the film thickness and another characteristic length $L$ for the corresponding 2D porous structure are combined to modify the bulk phonon MFPs with the Matthiessen rule, i.e., $\Lambda_{eff} = \left(\frac{1}{\Lambda_{Bulk}} + \frac{1}{L} + \frac{2|cos\theta|}{h}\right)^{-1}$. Here $\theta$ is the included angle between the phonon traveling direction and the cross-plane direction. In its calculations, $k_L$ is integrated over all solid angles and angular frequencies, and then summed over all phonon branches.[61] Although the thickness variation can be easily addressed in $\Lambda_{eff}$, this model again assumes completely diffusive film-surface phonon scattering and is lack of the flexibility to treat the film surfaces differently from pore edges.

In comparison, existing simulations or calculations mainly focus on the cases with diffusive film top/bottom surfaces. In many cases, the calculations can be very complicated with varied film thicknesses and surface phonon specularity. This critical issue can be easily addressed in the proposed two-step model.

## 4. Summary

In summary, a simple two-step model has been developed to compute $k_L$ for rectangular nanowires and general nanoporous structure etched from a solid thin film. With the two-step phonon MFP modification, the phonon specularity difference for the original film surfaces and the etched sidewalls can be separately addressed. Using this model, the influence of the film thickness dependence can be easily considered in data analysis. When an analytical expression is not available, the required $L_{eff}$ for certain 2D patterns can always be determined by matching predictions of the phonon MC simulations and the analytical model. Then this $L_{eff}$ can be used for varied film thicknesses and specularity for the film surfaces. The model can be easily extended to the most complicated phonon transport studies of nanoporous thin films, where the exact phonon dispersion and wave-vector dependence of phonon MFPs are considered.[32] For thermal designs of such nanoporous thin films,[13, 62] the film thickness must be considered and the proposed model can play an important role in such applications.

**Declaration of Competing Interest**

The authors declare that there are no conflicts of interest.

**Acknowledgements**

We acknowledge the support from the U.S. Air Force Office of Scientific Research (award number FA9550-16-1-0025) for studies on nanoporous materials and National Science Foundation (grant number CBET-1803931) for phonon MC simulations. An allocation of computer time from the UA Research Computing High Performance Computing (HPC), High Throughput Computing (HTC) at the University of Arizona is gratefully acknowledged.

References

1. A. M. Marconnet, M. Asheghi and K. E. Goodson, *Journal of Heat Transfer*, 2013, **135**, 061601.

2. D. G. Cahill, P. V. Braun, G. Chen, D. R. Clarke, S. Fan, K. E. Goodson, P. Keblinski, W. P. King, G. D. Mahan, A. Majumdar, H. J. Maris, S. R. Phillpot, E. Pop and L. Shi, *Applied Physics Reviews*, 2014, **1**, 011305.
3. Q. Hao, H. Zhao, Y. Xiao and M. B. Kronenfeld, *International Journal of Heat and Mass Transfer*, 2018, **116**, 496-506.
4. Q. Hao, H. Zhao and Y. Xiao, *Journal of Applied Physics*, 2017, **121**, 204501.
5. D. Song and G. Chen, *Applied Physics Letters*, 2004, **84**, 687-689.
6. J. Lee, W. Lee, G. Wehmeyer, S. Dhuey, D. L. Olynick, S. Cabrini, C. Dames, J. J. Urban and P. Yang, *Nature Communications*, 2017, **8**, 14054.
7. J. Lim, H.-T. Wang, J. Tang, S. C. Andrews, H. So, J. Lee, D. H. Lee, T. P. Russell and P. Yang, *ACS Nano*, 2016, **10**, 124-132.
8. J. Tang, H.-T. Wang, D. H. Lee, M. Fardy, Z. Huo, T. P. Russell and P. Yang, *Nano Letters*, 2010, **10**, 4279-4283.
9. J.-K. Yu, S. Mitrovic, D. Tham, J. Varghese and J. R. Heath, *Nature Nanotechnology*, 2010, **5**, 718-721.
10. R. Yanagisawa, J. Maire, A. Ramiere, R. Anufriev and M. Nomura, *Applied Physics Letters*, 2017, **110**, 133108.
11. J. Maire, R. Anufriev, R. Yanagisawa, A. Ramiere, S. Volz and M. Nomura, *Science Advances*, 2017, **3**, e1700027.
12. M. R. Wagner, B. Graczykowski, J. S. Reparaz, A. El Sachat, M. Sledzinska, F. Alzina and C. M. Sotomayor Torres, *Nano Letters*, 2016, **16**, 5661-5668.
13. G. Romano and J. C. Grossman, *Applied Physics Letters*, 2014, **105**, 033116.
14. Q. Hao, Y. Xiao and Q. Chen, *Materials Today Physics* 2019, **10**, 100126.
15. W. Park, J. Sohn, G. Romano, T. Kodama, A. Sood, J. S. Katz, B. S. Y. Kim, H. So, E. C. Ahn, M. Asheghi, A. M. Kolpak and K. E. Goodson, *Nanoscale*, 2018, **10**, 11117-11122.
16. K. Hippalgaonkar, B. Huang, R. Chen, K. Sawyer, P. Ercius and A. Majumdar, *Nano letters*, 2010, **10**, 4341-4348.
17. D. Li, Y. Wu, P. Kim, L. Shi, P. Yang and A. Majumdar, *Applied Physics Letters*, 2003, **83**, 2934-2936.
18. N. K. Ravichandran and A. J. Minnich, *Physical Review B*, 2014, **89**, 205432.
19. Q. Hao, D. Xu, H. Zhao, Y. Xiao and F. J. Medina, *Scientific reports*, 2018, **8**, 9056.
20. Y. Xiao, Q. Chen, D. Ma, N. Yang and Q. Hao, *ES Materials & Manufacturing*, 2019, **5**, 2–18.
21. G. Chen, *Nanoscale Energy Transport and Conversion: A Parallel Treatment of Electrons, Molecules, Phonons, and Photons*, Oxford University Press, New York, 2005.
22. J. M. Ziman, *Electrons and phonons: the theory of transport phenomena in solids*, Oxford University Press, 2001.
23. Z. M. Zhang, *Nano/microscale heat transfer*, McGraw-Hill New York, 2007.
24. N. K. Ravichandran, H. Zhang and A. J. Minnich, *Physical Review X*, 2018, **8**, 041004.
25. D. Gelda, M. G. Ghossoub, K. Valavala, J. Ma, M. C. Rajagopal and S. Sinha, *Physical Review B*, 2018, **97**, 045429.
26. J. S. Heron, T. Fournier, N. Mingo and O. Bourgeois, *Nano Letters*, 2009, **9**, 1861-1865.
27. X. Wang and B. Huang, *Scientific reports*, 2014, **4**.
28. C. Shao, Q. Rong, M. Hu and H. Bao, *Journal of Applied Physics*, 2017, **122**, 155104.
29. C. Shao, Q. Rong, N. Li and H. Bao, *Physical Review B*, 2018, **98**, 155418.
30. X. Wang and B. Huang, *Scientific Reports*, 2014, **4**, 6399.

31. F. Kargar, B. Debnath, J.-P. Kakko, A. Säynätjoki, H. Lipsanen, D. L. Nika, R. K. Lake and A. A. Balandin, *Nature Communications*, 2016, **7**, 13400.
32. A. Jain, Y.-J. Yu and A. J. McGaughey, *Physical Review B*, 2013, **87**, 195301.
33. Q. Hao, Y. Xiao and H. Zhao, *Journal of Applied Physics*, 2016, **120**, 065101.
34. Z. Hashin and S. Shtrikman, *Journal of applied Physics*, 1962, **33**, 3125-3131.
35. Q. Hao, G. Chen and M.-S. Jeng, *Journal of Applied Physics*, 2009, **106**, 114321/114321-114310.
36. J.-P. M. Péraud and N. G. Hadjiconstantinou, *Physical Review B*, 2011, **84**, 205331.
37. Z. Wang, J. E. Alaniz, W. Jang, J. E. Garay and C. Dames, *Nano letters*, 2011, **11**, 2206-2213.
38. J. Maire, R. Anufriev, T. Hori, J. Shiomi, S. Volz and M. Nomura, *Scientific Reports*, 2018, **8**, 4452.
39. S. Gluchko, R. Anufriev, R. Yanagisawa, S. Volz and M. Nomura, *Applied Physics Letters*, 2019, **114**, 023102.
40. Y. Kage, H. Hagino, R. Yanagisawa, J. Maire, K. Miyazaki and M. Nomura, *Japanese Journal of Applied Physics*, 2016, **55**, 085201.
41. H. Casimir, *Physica*, 1938, **5**, 495-500.
42. A. K. McCurdy, H. J. Maris and C. Elbaum, *Physical Review B*, 1970, **2**, 4077-4083.
43. J. Heron, T. Fournier, N. Mingo and O. Bourgeois, *Nano letters*, 2009, **9**, 1861-1865.
44. X. Lü, J. Chu and W. Shen, *Journal of applied physics*, 2003, **93**, 1219-1229.
45. W. Park, D. D. Shin, S. J. Kim, J. S. Katz, J. Park, C. H. Ahn, T. Kodama, M. Asheghi, T. W. Kenny and K. E. Goodson, *Applied Physics Letters*, 2017, **110**, 213102.
46. W. Liu, Carnegie Mellon University, 2005.
47. W. Maszara, *Journal of the Electrochemical Society*, 1991, **138**, 341-347.
48. C. Dames, S. Chen, C. T. Harris, J. Y. Huang, Z. F. Ren, M. S. Dresselhaus and G. Chen, *Review of Scientific Instruments*, 2007, **78**, 104903-104913.
49. L. Shi, *Appl. Phys. Lett.*, 2008, **92**, 206103.
50. K. Esfarjani, G. Chen and H. T. Stokes, *Physical Review B*, 2011, **84**, 085204.
51. P. E. Hopkins, P. T. Rakich, R. H. Olsson, I. F. El-Kady and L. M. Phinney, *Applied Physics Letters*, 2009, **95**, 161902.
52. P. E. Hopkins, L. M. Phinney, P. T. Rakich, R. H. Olsson and I. El-Kady, *Applied Physics A*, 2011, **103**, 575-579.
53. S. Alaie, D. F. Goettler, M. Su, Z. C. Leseman, C. M. Reinke and I. El-Kady, *Nature Communications*, 2015, **6**, 7228.
54. Y.-C. Hua and B.-Y. Cao, *The Journal of Physical Chemistry C*, 2017, **121**, 5293-5301.
55. Q. Hao, Y. Xiao and H. Zhao, *Applied Thermal Engineering*, 2017, **111**, 1409-1416.
56. K. D. Parrish, J. R. Abel, A. Jain, J. A. Malen and A. J. McGaughey, *Journal of Applied Physics*, 2017, **122**, 125101.
57. A. J. McGaughey and A. Jain, *Applied Physics Letters*, 2012, **100**, 061911.
58. A. M. Marconnet, T. Kodama, M. Asheghi and K. E. Goodson, *Nanoscale and Microscale Thermophysical Engineering*, 2012, **16**, 199-219.
59. G. Romano and J. C. Grossman, *Journal of Heat Transfer*, 2015, **137**, 071302.
60. G. Romano and A. M. Kolpak, *Applied Physics Letters*, 2017, **110**, 093104.
61. C. Huang, X. Zhao, K. Regner and R. Yang, *Physica E: Low-dimensional Systems and Nanostructures*, 2018, **97**, 277-281.
62. Z. Yu, L. Ferrer-Argemi and J. Lee, *Journal of applied physics*, 2017, **122**, 244305.